\documentstyle[a4p,epsfig,cite,12pt]{article}
\begin{document}

\def\z{${\rm Z}^0$}
\def\ep{e$^+$e$^-$}
\def\ezh{{{\ep} $\to$ {\z} $\to {hadrons}$}}
\def\ezhc{{{\ep}$\to${\z} $\to{hadrons}$}}
\def\eh{{{\ep} $\to {hadrons}$}}
\def\ie{{\sl i.e.}}
\def\eg{{\sl e.g.}}
\def\ea{{\sl et al.}}
\def\vrs{{\sl vs.}}
\def\col{Collab.}
\def\JT{{\sc Jetset}}
\def\OP{OPAL}
\def\DE{DELPHI}

\def\al{\langle}
\def\ar{\rangle}
\def\ka{\kappa}

\def\vs{\vspace*}
\def\hs{\hspace*}
\def\bea{\begin{eqnarray}}
\def\eea{\end{eqnarray}}
\def\be{\begin{equation}}
\def\ee{\end{equation}}
\def\la{\label}
\def\bga{\left( \begin{array}}
\def\ena{\end{array} \right)}
\def\nwp{\newpage}
\def\ct{\cite}
\def\bi{\bibitem}
\def\ni{\noindent}
\def\fn{\footnote}

\def\pT{p_T}
\def\phi{\Phi}
\def\yf{y$$\times$$\phi}
\def\yf{y$$\times$$\phi}
\def\yp{y$$\times$$\pT}
\def\fp{\phi$$\times$$\pT}
\def\3d{y$$\times$$\phi$$\times$$\pT}
\def\lsim{\:{\stackrel{<}{_\sim}}\:}

\date{}

\def\jour#1#2#3#4{{#1} {#2} (19#3) #4}
\def\jourpt#1#2#3{{#1} (19#2) #3}
\def\jourm#1#2#3#4{{#1} {#2} (20#3) #4}
\def\jourm#1#2#3#4{{#1} {#2} (20#3) #4}
\def\PRp{Phys. Reports}
\def\PRD{Phys. Rev. {D}}
\def\PRC{Phys. Rev. {C}}
\def\PRL{Phys. Rev. Lett. }
\def\AP{Acta Phys. Pol. {B}}
\def\ZP{Z. Phys.  {C}}
\def\EPJ{Eur. Phys. J. {C}}
\def\IJ{Int. J. Mod. Phys. {A}}
\def\NIM{Nucl. Instr. Meth. {A}}
\def\CP{Comp. Phys. Comm.}
\def\TPS{Theor. Phys. Suppl.}
\def\PL{Phys. Lett.  {B}}
\def\NPA{Nucl. Phys.  {A}}
\def\NPB{Nucl. Phys.  {B}}
\def\MPL{Mod. Phys. Lett. {A}}
\def\JP{J. Phys. {G}}
\def\NC{Nuovo Cim.}
\def\UFN{Physics-Uspekhi}
\def\HEPC{High En. Phys. \& Nucl. Phys.}

     %%%%%%%%%%%%%%%%%%%%%%%%%%%%%%%%%%%%%%%%%%%%% 
     %                                           %        
     %          T I T L E   L I S T              %
     %                                           % 
     %%%%%%%%%%%%%%%%%%%%%%%%%%%%%%%%%%%%%%%%%%%%%

\renewcommand{\Huge}{\huge}
\textheight=24.5cm
\pagestyle{empty}
\hs{4.cm}

\vs{1cm}
\begin{center}

{\Large \bf
The effect of many sources\\
on the genuine multiparticle correlations\\ 
}
\bigskip
\bigskip
{\large Gideon Alexander\fn{Email address: alex@lep1.tau.ac.il}
and 
Edward K.G. Sarkisyan\fn{ 
        Email address: edward@lep1.tau.ac.il}
}\\ 
\medskip
{\small \it School of Physics and Astronomy,\\
The Raymond and Beverly Sackler Faculty of Exact Sciences,\\
Tel-Aviv University, IL-69978 Tel-Aviv, Israel
}
\vs{2.8cm}
\end{center} 
\medskip

\vs{-11.cm}
\begin{flushright}
{\bf OPAL-IP85}\\
{\bf TAUP 2630-2000}\\
{\bf hep-ph/0005212}\\
\end{flushright}
\vs{8.cm}

     %%%%%%%%%%%%%%%%%%%%%%%%%%%%%%%%%%%%%%%%%%%%% 
     %                                           %        
     %             A B S T R A C T               %
     %                                           % 
     %%%%%%%%%%%%%%%%%%%%%%%%%%%%%%%%%%%%%%%%%%%%%

\begin{abstract}
\vs{.7cm}
\ni
  We report on a study aimed to explore the dependence of the
genuine multiparticle correlations on the number of sources when the
influence of other possible factors during multihadron production is
avoided. The analysis utilised the normalised 
cumulants calculated
in three-dimensional phase space of the reaction {\ezhc} using a 
large Monte Carlo sample.
 The multi-sources reactions were simulated by overlaying a few independent
single {\ep} annihilation events.
 It was found that as the number of sources $S$ increases, the cumulants do
not change significantly their structure, but those of an order $q > 2$
decrease fast in their magnitude.
 This reduction in the one-dimensional
rapidity cumulants can be understood in terms
of combinatorial considerations of source mixing which dilutes 
the correlations by a factor of about $1/S^{q-1}$.
 The diminishing of the genuine correlations is consistent with recent
cumulant measurements in hadron and nucleus induced reactions and
should also be relevant to other dynamical correlations
like the Bose-Einstein one, in e$^+$e$^-\to$W$^+$W$^-$
$\to hadrons$ and in nucleus-nucleus reactions.  
 \end{abstract}

    %%%%%%%%%%%%%%%%%%%%%%%%%%%%%%%%%%%%%%%%%%%%%
    %                                           %
    %            PACS  &  KEYWORDS              %
    %                                           %
    %%%%%%%%%%%%%%%%%%%%%%%%%%%%%%%%%%%%%%%%%%%%%

\vs{2.cm}
{
\ni

\ni
}
\nwp
\thispagestyle{empty}

     %%%%%%%%%%%%%%%%%%%%%%%%%%%%%%%%%%%%%%%%%%%%% 
     %                                           %        
     %             M A I N    T E X T            %
     %                                           % 
     %%%%%%%%%%%%%%%%%%%%%%%%%%%%%%%%%%%%%%%%%%%%%

\pagestyle{plain}
\setcounter{page}{1}

\section{Introduction}

 During the last decade an increase interest has been shown for
the genuine multiparticle correlations in multihadron final states of
hadronic, {\ep} and other reactions \ct{revi}.
 Recently OPAL, in its study of {\ep} annihilations on the Z$^0$ mass, has
established the existence of strong genuine multihadron correlations up to
the fifth order \ct{Ogc}.
 In hadron-hadron, like proton-proton, collisions the correlations of more
than three particles have also been observed \ct{hAc,na22,hhc}.
 In contrast to this situation, in heavy ion collisions, at low energies
and/or in reactions of light nuclei, genuine correlations are found to
have non-zero values only up to the third order \ct{AAc2}. 
 Furthermore it has been found out that in general these correlations
become weaker as the reaction average multiplicity increases.
 In nucleus-nucleus collisions at high energies, of tens and hundreds
GeV per nucleon, the two-particle correlations are the only one that
survive \ct{hAc,nfcd,isn,AAc1}.
 \\

\ni
 This correlation dependence on the average multiplicity is very similar
to the one observed in the investigations of multiparticle dynamical
fluctuations, i.e. variation of many particle bunches in restricted phase
space regions \ct{revi}.
 In these studies, known as intermittency analyses, the average
multiplicity dependence has been proposed to be the consequence of a
mixing of several independent emission sources \ct{is1,is2,isn}.
 As a result, the dynamical fluctuations in nucleus-nucleus collisions are
already well accounted for by two-particle correlations \ct{nfcd,ncf},
whereas in hadron-hadron interactions \ct{hhcf,na22} and in {\ep}
annihilations \ct{Ogc} higher order genuine correlations do exist.
 \\

\ni
 An analogous situation is also observed in Bose-Einstein correlations
(BEC) where identical bosons are correlated when they emerge from
the interaction in nearby phase space.
 A genuine three-pion BEC has been detected in hadron-hadron reactions
\ct{hhbec} and found to be even more pronounced in {\ep} annihilations
\ct{eebec}.
 In contrast to these, no genuine three-pion BEC were found in
nucleus-nucleus collisions where the three-body correlations were 
well reproduced in terms of two-particle BEC \ct{AAbec}.
 Since the intermittency phenomenon and BEC seem to be closely related
\ct{revi}, the dependence of many sources on the strength of the BEC
cannot be excluded.
 The superposition of emitters may also be a reason for the suppression of
BEC of hadrons produced from W-boson pairs in {\ep} annihilations at LEP2
energies where the overlapping of hadrons affect the accuracy of the W
mass measurements \ct{WWbec}.
 \\

\ni
 All this, as well as the obvious intrinsic interest in the genuine
correlations which carry most of the dynamics of the hadron production
process, points to the need of dedicated studies aimed to investigate the
correlation dependence on the number of emission sources.
 Here we propose to study this dependence by grouping several {\eh}
events to represent a multi-emission sources of particles.
 To obtain significant results, even when only few sources are
considered, one needs a very high statistics, like that which can be
supplied by a Monte Carlo (MC) generated sample, to minimise the
calculation error and thus be sensitive to the correlations dependence on
the number of sources.
 Another advantage of using MC generated events is in its possibility 
to generate a multihadron sample free from 
contamination of other processes like, for
example, {\ep} $\to \tau^+\tau^- \to hadrons$.
 \\  

\ni
 In this Letter we present the results of a MC study on the effect of
several emission sources on the genuine higher order multiparticle
correlations.
 The study was based on a generated sample of about 
$5 \times 10^6$ events of the
reaction {\ezh} which passed a full simulation of the OPAL detector at
LEP and did reproduce 
rather well the measured genuine high order correlations present in the
OPAL hadronic {\z} decay data \ct{Ogc}.
Moreover, the {\ezh} annihilations should    
represent well the  
one emission source situation in contrast to events produced in
hadron-hadron interactions. 
Our analysis on the dependence of many sources
on the genuine correlations was thus carried out in a way that avoided  
effects of other reaction features, like the multiplicity 
which was discussed recently in connection with 
two-particle BEC analyses in \ct{bec-is} and \ct{bec-iss}.
In as much that final state interactions
between hadrons coming from different sources
can be neglected,
our correlation study based on {\ezh} annihilations 
may be extended to other types of reactions since the
hadronisation process is believed to rest on a common basis \ct{qcd}.

\section{The analysis method}

 The analysis is based on a generated sample of hadronic Z$^0$ decays
using the {\JT} 7.4 MC program \ct{JT} including a full simulation of the
OPAL detector at LEP \ct{MCOd}.
 The MC sample also included initial-state radiation and effects of finite
lifetimes.
 The parameters of the program were tuned to yield a good description of
the measured event shape and single particle distributions \ct{MCO}.
 \\

\ni
 The selection criteria for multihadron events used here are identical to
the ones previously utilised by OPAL in their recent data analysis of
multiparticle correlations \ct{Ogc}.
 In particular, selected events were required to have at least five
charged tracks each having at least 20 measured
points in the jet chamber where the first point had
to be closer than 40 cm from the beam axis.
 The cosine of the polar angle of the event sphericity axis with respect
to the beam direction was required to be less than 0.7 to ensure the event
to be within the volume of the detector.
 The sphericity axis was calculated by using all accepted tracks and
electromagnetic and hadronic calorimeter clusters.
 \\ 

\ni
 To simulate several emission sources we did overlay several {\ezh}
generated events and analysed the correlations between pions as if they
were created in a single event.
 The kinematic variables are defined within each generated {\ep} event
with respect to its own sphericity axis.
 For correlation analyses of variables like rapidity this procedure is
equivalent to the one where the events are rotate to a common sphericity
axis.
 This simulates multi-sources' events with hadrons, here all taken to be
pions, emerging from a common emitter. To note is that
in this procedure 
the average event multiplicity is directly proportional
to the number of sources. In our analysis each generated event was
used only once, and hence required
a very large MC event statistics.
\\

\ni
 To extract the genuine dynamical $q$-particle correlations, we used
bin-averaged normalised factorial cumulant moments, or cumulants, first
proposed in Ref. \ct{cum} as a tool for search for genuine multiparticle
correlation,

\be
K_q={1\over M}
\sum_{m=1}^{M} \int_{\delta y} \prod_i {\rm d}y_i 
\frac 
 {C_q(y_1,\ldots, y_q)}
 {[ \int_{\delta y} {\rm d}y\rho_1(y)] ^q}\:,
\la{kmy}
\ee
\ni 
where $C_q(y_1,\ldots, y_q)$ are the $q$-particle correlation functions
given by the inclusive $q$-particle density distributions
$\rho_q(y_1,\ldots, y_q)$ in terms of cluster expansion, {\eg},

\be
C_3(y_1,y_2,y_3)=\rho_3(y_1,y_2,y_3)
    -\sum_{(3)}\rho_1(y_1)\rho_2(y_2,y_3)
    +2\,\rho_1(y_1)\rho_1(y_2)\rho_1(y_3)\:.
\la{cr}
\ee

\ni
 Here $M$ is the number of equal bins of a width $\delta y$ into which
the event phase-space is divided and the subscript (3) denotes the
number of permutations.
 For simplicity we show all formulae in one-dimensional ({\eg}, rapidity)
phase space.
\\

\ni
 The feature of the $C_q$-functions is that they vanish whenever there are
no genuine correlations, {\ie}, the correlations are due to those existing
in lower orders.
 The correlations extracted are of a dynamical nature since the cumulants
share with normalised factorial moments (the intermittency analysis tool)
the property of statistical noise suppression.
 \\

\ni
In this paper we computed the cumulants as they were used in experimental
studies, in particular we used the form applied in Ref. \ct{Ogc} namely,

\be 
K_q= 
\frac{
{\cal N}^q \cdot 
\sum_{m=1}^{M} k_q^{(m)}
}{ 
\sum_{m=1}^{M} N_m(N_m-1)\cdots(N_m-q+1)
}\:\;. 
\la{kmh}
\ee 
 Here, the factors $k_q^{(m)}$ are the unnormalised factorial cumulant
moments, or the Mueller moments \ct{math}, calculated for the $m$th bin.
 These factors represent the correlation functions $C_q$ integrated over
the bin and $N_m$ is the number of particles in the $m$th bin summed over
all the $\cal N$ events.
 The definition (\ref{kmh}) takes into account the non-uniform shape of
the single-particle distribution and the bias when the cumulants are
computed at small bins.
 \\

\ni
 The cumulant calculations were performed in the three-dimensional phase
space of the kinematic variables commonly utilised in this kind of studies
\ct{revi}, namely:
 \\

 \begin{itemize}
 \item The rapidity, $y=\ln \sqrt {(E+p_{\|})/(E-p_{\|}) }$, with $E$ and
$p_{\|}$ being the energy and longitudinal momentum of the hadron in the
interval $-2.0 \leq y\leq 2.0$;
 \item The transverse momentum in the interval $0.09 \leq \pT\leq 2.0$
GeV/$c$;
 \item The azimuthal angle, $0\leq\phi<2\pi$, calculated with respect to
the eigenvector of the momentum tensor having the smallest eigenvalue, in
the plane perpendicular to the sphericity axis.
 \end{itemize} 
 These variables are defined with respect to the sphericity axis, in a way
and within the intervals similar to those used in a recent OPAL analysis
\ct{Ogc} and in other cumulant studies \ct{revi}.

\section{Genuine correlations and the number of sources}

\subsection{Monte Carlo studies}

\ni
 In Fig. 1 we compare the MC based cumulants of orders $q=2,3$ and 4
calculated from a single {\ezh} source (solid symbols) with those
obtained by overlaying seven such events representing seven hadronic
sources (open symbols).
 The calculations were performed in one-dimensional rapidity, in
two-dimensional rapidity {\vrs} azimuthal angle subspaces and in
three-dimensional phase space of rapidity, azimuthal angle and transverse
momentum.
 \\

\ni
 The following observations can clearly be made from Fig. 1.

 \begin{itemize} 
\item 
 The existence of a dynamical component, {\ie} rise of the cumulants with
increasing number of bins $M$, is seen to be present both in the single
source as well as in the case of many sources.
 Although the slopes of this scaling behaviour are smaller for several
sources than for a single source, they are still strongly present.         
 It is also evident that the character of the scaling is kept as the
number of sources increases {\eg}, for one source as well as for seven
sources the one-dimensional (rapidity) 
cumulants level off at the same $M$ value. No such saturation exists
for the one and seven sources cumulants in the two and three dimensions.     

\item 
The genuine dynamical correlations, measured by the cumulants,
significantly decrease with the increase of the number of sources.
 This decrease is stronger for higher order correlations namely, whereas
the two-particle cumulants suffer a reduction of an order of magnitude
as the number of sources increases from one to seven, the four-particle
cumulants diminish by three or four orders of magnitude.

\item 
 The hierarchy of the $K_q$ cumulants is reversed as the number of sources
increases.
  The cumulants derived from the single-source events increase with
increasing $q$-order so that $K_2^{(1)} < K_3^{(1)} < K_4^{(1)}$, whereas
the hierarchy in
the cumulants calculated for seven sources is reversed
namely, $K_2^{(7)} > K_3^{(7)} \geq K_4^{(7)}$.
 In addition, the multi-sources cumulants of order $q>2$ have almost the
same reduced value namely, $K_3^{(7)} \approx K_4^{(7)} \lsim {\cal
O}(0.1)$. 
 This last feature does not change as the dimension increases.

\item 
 The overall dominant feature of the analysis results is the diminishing
value of the higher order cumulants as the sources number increases
leaving the $K_2$ to be  the dominant genuine multiparticle
correlations.
 \end{itemize}

\ni
 The observed diminished correlations is further illustrated in Fig. 2
where the $M$-averaged one dimensional (rapidity)  
$K_q^{\rm av}$ cumulants, 
 are plotted
with their errors against the number of sources.
 The cumulants were averaged over the $M\ge10$ region where they are seen
in Fig. 1 to approach an almost constant value.
 The values of these $K_q^{\rm av}$ are also listed in Table 1 together
with the OPAL measured data cumulants \ct{Ogc} of single {\ezh} events.
These data cumulants for $q\leq3$ are seen to agree with those derived
from the MC sample. 
 The one source MC based $q\ =\ 4$ cumulant lies lower than the measured
data value but is still consistent within errors.
 \\ 
    
\ni
 It is obvious from Fig. 2 that the $M$-averaged rapidity cumulants of order
$q>2$ decrease fast with the increase of the number of sources.
 Already for two sources the hierarchy changes and the two-particle
correlations visibly dominate over the higher order ones.
 At higher number of sources the dominant role of the two-particle
correlations is even more pronounced.
 \\

\subsection{Correlation dilution due to source mixing}

\ni
 Within the procedure adopted here for the simulation of multi-source
events, it is clear that if a genuine correlation exists it can only be
detected in groups of $q$ pions emerging from the very same source.
 In those $q$-group combinations which emerge from at least two sources,
genuine correlations should not be present.
 This means that for $K_q$ cumulants that are calculated over all possible
$q$-pion groups, the higher the number of sources the more diluted will
be the signal for genuine correlations.
 \\

\ni
For a given $q$-order the genuine correlation dilution factor is thus:

\be
R_q\ =\ \frac{P_q^{\rm G}}{(P_q^{\rm G} +P_q^{\rm NG})}\ ,
\la{dif}
\ee        
 where $P_q^{\rm G}$ denotes the number of $q-$particle groups, e.g.,
pairs or triplets of pions, which emerge from the same source.
 The term $P_q^{\rm NG}$ stands for the number of all possible
combinations of $q-$particle groups which emerge from at least two
sources.
 \\

\ni
 Since all sources are produced in the same reaction and at the same
energy, they do have an identical average charged multiplicity.
 For the estimation of $R_q$ we assume that all the $S$ sources have 
the same fixed charged multiplicity $n$. 
 In this case one has $P_q^{\rm G} = S\, {n \choose q}$, and the dilution
factors at $q=2$, 3 and 4 are given by

\bea
\nonumber
R_2& = & \frac{ {n \choose 2}\,S}
                {{n \choose 2}\, S+n^2\,{S \choose2}
              }\ \
{\stackrel{n\gg 1}{\longrightarrow}}\ \ \frac{1}{S}  \:\, ,
\\ \nonumber 
\\ \label{difn} 
R_3& = & \frac{ {n \choose 3}\,S}
    {{n \choose 3}\,S+n^3\, {S\choose 3}+2\,n\,{n\choose 2}{S\choose 2}
                } \ \
{\stackrel{n\gg 2}{\longrightarrow}}\  \ \frac{1}{S\,^2} \:\,,
\\ \nonumber
\\ \nonumber
R_4& = & \frac{ {n \choose 4}\,S}
                {{n \choose 4}\,S +
     n^4\,{S\choose 4}+3\,n^2\,{n\choose 2}{S\choose 3} +
     {n\choose 2}^2\ {S\choose 2}+ 2\,n\,{n\choose 3}{S\choose 2}}\ \ \
{\stackrel{n\gg 3}{\longrightarrow}}\ \ \frac{1}{S\,^3} \:\,,
\nonumber 
\eea 
 where the denominators include the number of all possible $q$-particle
combinations in $S$ sources of charged multiplicity $n$.
  This dilution factor dependence on the number of sources can also be
derived in terms of cumulants \ct{is1}. 
 \\

\ni
 From these $R_q$ relations one can show that as long as $n \gg q$ one
obtains a general expression for the dilution factor,

\be
R_q\  \ {\stackrel{n\gg q}{\longrightarrow}}\ \ 
            \frac{1}{S\,^{q-1}}\:.
\label{difi}
\ee   
\ni 
 To compare the dilution factors $R_q$ with our correlation results shown
in Fig. 2, they do have to be multiplied by $K_q^{{\rm av}(1)}$ which is a
measure of the genuine $q$-order correlation present in a single sources.
 The solid lines shown in Fig. 2 thus represent the
diluted cumulants $K_q^{\rm av}=K_q^{{\rm av}(1)}\times R_q$. 
 The striped areas in which the lines are embedded are the allowed regions
when $q$ is not neglected with respect to the multiplicity $n$.
 The agreement between the cumulant calculations and the dilution factors
predictions is really remarkable for $q=2$ and $3$ and certainly is 
still well within the rather large errors of the $q = 4$ cumulants.
 \\

\ni
 For the order $q\ =\ 2$ one can relax the fixed charged multiplicity
assumption and allow them to be different and still retain
the $R_q \simeq 1/S^{q-1}$ relation as long as the multiplicity
distribution is of a Poisson nature. This however is not the case
for orders higher than 2. Nevertheless for order 
$q\ =\ 3$ the relation $R_3 = 1/S^2$, derived from the fixed
multiplicity assumption, is still valid as it describes
well the      
$K^{\rm av}_3$ values up to at least thirteen sources (see Fig. 2).
The large cumulants' errors associated with the $q\ =\ 4$ order
prohibits to judge how accurate is the $R_4\ =\ 1/S^3$ relation.\\
 
\ni
 An additional interesting and useful application of the relation $R_q
\simeq 1/S^{q-1}$ is that it offers a method to estimate the average
number
of sources $\al S\ar$ via the cumulant averaged values over the large $M$
region of two sequential $q$-orders through the ratio,

\be
\al S\ar\ \simeq \ 
\frac{K_{q+1}^{{\rm av(1)}}}{K_q^{{\rm av(1)}}}      
\times 
\frac{K_q^{{\rm av}}}{K_{q+1}^{{\rm av}}} \ .
\label{useful}
\ee   

\subsection{Comparison with hadron and nucleus induced reactions}

\ni
 As is already mentioned in the introduction, the genuine correlations
measured in {\ep} annihilations \ct{Ogc} are found to be weaker in
hadronic interactions \ct{hhcf,hAc,na22,hhc} and even more so in nuclear
collisions \ct{hAc,nfcd,isn,AAc1,AAc2}.
 In nucleus-nucleus collisions at ultra-relativistic energies only the
second-order correlations were so far detected \ct{hAc,nfcd,isn,AAc1}.
 \\

\ni
 In Table 2 we list the results obtained by several experiments
\ct{hAc,AAc1,isn,AAc2} on the $M$-averaged rapidity cumulant
values for $q=2$ and 3.
 These average values were taken over the $M$-region where the cumulants
are seen in the published figures to reach a constant level.
 The hadronic reactions and their cumulants values are ordered according
to their reported mean charged multiplicity, from the lowest value to the
highest one.
 \\

\ni
 Table 2 shows that in hadron including nucleus induced 
reactions the two-particle
correlations decrease rather fast as the mean multiplicity increases.
 However the three-particle correlations are found to be essentially
non-existing even at moderately small mean multiplicity.
 Notwithstanding the possibility that production of hadrons in {\ep}
annihilation may well be simpler than in hadron induced reactions, it may
nevertheless be instructive to relate our findings to the measured
correlation data listed in Table 2.
 In as much that the mean multiplicity increases with the number of
sources, the decrease in the two-particle correlations and the absence of
three-pion correlations in nucleus induced reactions is consistent
with our findings which demonstrated the dilution of the correlations with
increased number of sources.
 A quantitative comparison between our findings and the correlations in
nucleus-nucleus and hadron induced reactions is hard mainly because of the
large errors associated with the average cumulant values. 
 In particular the application of relation (\ref{useful}) is prohibited
because most of the $K^{\rm av}_3$ are consistent within errors with zero.
 \\

\ni
 Recently the two-pion BEC have been studied \ct{bec-iss} in ${\bar {\rm
p}}$p reaction at centre of mass energy of 630 GeV as a function of
multiplicity by using the normalised cumulants method similar to the one
used here.
 In that analysis it has been found that the correlations 
of the cumulants of the like-sign pions as well as the opposite-sign pions 
decrease with the multiplicity $n_{ch}$. 
 From our analysis we expect the pair correlation to decrease as $1/S$,
where $S$ is the number of sources. This indicates that indeed the
multiplicity is at least partially proportional to $S$.
 The BEC dependence has also been investigated in the framework of the
totally coherent emission picture \ct{bec-coh} and in the
quantum optical approach \cite{bec-is} where the conclusions were that
these correlations are weaker as the multiplicity increases.
 \\ 

\section{Summary and conclusions}
\label{sum}

\ni
To investigate the effect of many emission sources on the genuine
correlations in mutihadron final state we adopted a procedure which
should minimise the confusion introduced by other variables
like charged multiplicity. 
 For the genuine correlations
measurement we utilised the normalised cumulant method.
To simulate the situation
of many sources event we did overlay Monte Carlo generated hadronic
Z$^0$ events treating them as one event. This
\JT 7.4 MC
sample of some five million events,  
tuned to the OPAL data taken at LEP1 on the Z$^o$ mass, has
previously described rather well the measured correlations
in the {\ezh} data.\\
 
\ni
The results obtained here  
show that the cumulants, obtained from a single-source events
and from events of many sources, almost do not change their structure with
the decrease of the width of phase-space bins. This means that the
scaling is preserved  although larger
slopes are seen to be in the case of one source compared with those for
several sources.
However, when the number of sources increases, the cumulants of order
higher than two are suppressed and diminish to zero due to source 
mixing.
 The two-particle cumulants are also somewhat reduced in 
their value but they are way above the higher order ones and they
are seen to completely dominate when the source number $S$ exceeds
the value ten.\\

\ni
The one-dimensional (rapidity)
correlations are very well reproduced by assuming that
genuine correlation of the order $q$ can only be present when
all the $q$ hadrons are emerging from the same source. Therefore
the dilution of the genuine correlation signal is proportional to the
ratio of the probability that the $q$ hadrons will come from the same
source. From simple combinatorial considerations this probability
is approximately equal to $1/S
^{q-1}$. Thus a measurement of the 
one-dimensional
correlation for two sequential orders renders the number of sources.\\

\ni
The genuine correlations measured in hadron and nucleus 
induced reactions do follow
qualitatively the findings of our work. In particular in
nucleus-nucleus
reactions, where many sources are expected to contribute to the final
hadronic state, the
$q > 2$ orders are very small and indeed consistent with zero.
The $q=2$ order is still present but it is also getting smaller as 
the atomic number of the nuclei increases. 
 The general decrease of the second order cumulants with the increase of
multiplicity re-confirms the belief that the higher the multiplicity the
larger the number of sources.
 Our results may also be useful for the understanding of other types of
measured correlations like the Bose-Einstein interference of two and more
identical bosons.  
 It has been previously pointed out \ct{bec-w} that in the absence of
final state interactions the BEC of the e$^+$e$^- \to$
W$^+$W$^- \to hadrons$ will be half of that of the {\z} decay to hadrons. 
 From our study it follows that the two-particle BEC, or any other
correlations, in the two-source reaction e$^+$e$^- \to$ W$^+$W$^- \to
hadrons$ should be reduced by a factor two as compared to that of
the hadrons emerging from one W-boson.

\section*{\small\bf Acknowledgements}

\ni
We would like to thank our colleagues from the {\OP} Collaboration 
for allowing us to use the Monte Carlo sample of
Z$^o$ decays into hadrons.
Special thanks are also due G. Bella, S. Kananov and S. Nussinov for many
helpful and inspiring discussions.

%\nwp

     %%%%%%%%%%%%%%%%%%%%%%%%%%%%%%%%%%%%%%%%%%%%% 
     %                                           %        
     %          R E F E R E N C E S              %
     %                                           % 
     %%%%%%%%%%%%%%%%%%%%%%%%%%%%%%%%%%%%%%%%%%%%%

\nwp

     %%%%%%%%%%%%%%%%%%%%%%%%%%%%%%%%%%%%%%%%%%%%%
     %                                           %
     %                T A B L E S                %
     %                                           %
     %%%%%%%%%%%%%%%%%%%%%%%%%%%%%%%%%%%%%%%%%%%%%

\vspace{0.3cm}
\begin{table}
\caption{
 The Monte Carlo $M$ averaged rapidity $K_q^{\rm av}$ cumulants 
obtained in the {\ezh}
reaction compared with those obtained in a recent OPAL 
measurement of single
data events.
 The averages were taken over the $M$ region
where the cumulants approached a constant value (see Fig. 1).
 }
\begin{center}
\smallskip
\begin{tabular}{ccccc} \hline  \hline
{\small No. of} & \multicolumn{3}{c}{$K_q^{\rm av}\,
   (M\geq10)$}\\\cline{2-4} 
{\small sources} & $q=2$ & $q=3$& $q=4$ & 
   {\raisebox{1.3ex}[0cm][0cm]{Sample}} \\ \hline
 & $0.45\pm 0.01$ & $ 0.67\pm 0.04$ & $1.36\pm 0.21$& Data \ct{Ogc}\\
\hline
1 & $0.486\pm 0.002$ & $0.632\pm 0.017$ & $0.950 \pm 0.225$ & MC\\
2 & $0.260\pm 0.001$ & $0.171\pm 0.006$ & $0.147 \pm 0.034$ & ''\\
4 & $0.124\pm 0.001$ & $0.041\pm 0.003$ & $0.016\pm 0.012$ & ''\\
7& $0.072\pm 0.001$ & $0.013\pm 0.003$ & $0.003\pm 0.008$ & ''\\
13\ \ & $0.037\pm 0.001$ & $0.004\pm 0.003$& $0.001\pm0.009$  & ''\\
\hline
\hline
\end{tabular}
\end{center}
\end{table}
\vspace*{.7cm}
\begin{table}
\caption{
 The $M$ averaged rapidity cumulants $K_q^{\rm av},\:$ 
of orders $q=2$ and 3  measured in several hadronic reactions.
 The cumulant values were averaged over the $M$ 
regions where they were seen to approach a constant value.
  These quoted values were estimated 
from the relevant published figures given in  
the references listed in the table.
 }
\begin{center} 
\smallskip
\begin{tabular}{cccllll} 
\hline 
\hline
{\raisebox{-1.3ex}[0cm][0cm]{$\al n_{\rm ch}\ar $}}
   & \multicolumn{2}{c}{$K_q^{\rm av}$} &
   {\raisebox{-1.3ex}[0cm][0cm]{Reaction}} & Beam energy & 
   {\raisebox{-1.3ex}[0cm][0cm]{Ref.}} \\\cline{2-3}
   & $q=2$ & $q=3$ &     & (GeV) & \\ \hline 
$\sim 8\ \ \ \ \ \ $ & $0.32\pm 0.02$  &  $0.26\pm 0.12$  & $\pi$p  & 250
&
    \ct{hAc}$^*$\\
21.1 & $0.21\pm 0.04$  &  $0.14\pm 0.18$  & pEm     & 200 &
   \ct{hAc}$^*$\\
$>50\ \ \ \ \ \ $ & $0.34\pm 0.02$  &  $0.12\pm 0.03$  & AuEm & 10.6$A$ &  \ct{AAc2}\\
73.3 & $0.21\pm 0.03$  &  $0.05\pm 0.07$  & SiEm   & 14.5$A$ &
   \ct{AAc1,AAc2}\\
81.1 & $0.20\pm 0.05$  &  $0.00\pm 0.12$  & OEm    & 60$A$ &  \ct{AAc1}\\
154.9 & $0.11\pm 0.05$  &  $0.01\pm 0.18$  & OEm     & 200$A$ &
  \ct{hAc}$^*$\\
216.1 & $0.25\pm 0.05$  &  $0.08\pm 0.10$  & SEm    & 200$A$ &
  \ct{AAc1}\\
272.6 & $0.09\pm 0.05$  &  $0.00\pm 0.18$  & SEm    & 200$A$ &
  \ct{hAc}$^*$\\
289.8 & $0.11\pm 0.08$  &  $0.02\pm 0.10$  & SAu & 200$A$ &
  \ct{isn}$^{**}$\\
355.0 & $0.08\pm 0.05$  &  $0.00\pm 0.08$  & SAu  & 200$A$ &
  \ct{hAc}$^*$\\
383.9 & $0.07\pm 0.04$  &  $0.00\pm 0.05$  & SAu  & 200$A$ &
 \ct{isn}$^{***}$\\
\hline
\hline
\end{tabular}
\end{center}
\vspace*{-1.95cm}
\end{table}
\ni
\hspace*{2.3cm}
$^*$ {\small Calculations are based on the measured factorial moments.}\\
\hspace*{2.3cm}
{$^{**}$ {\small Semicentral collisions.}\\
\hspace*{2.3cm}
{$^{***}$ {\small Central collisions.}\\

     %%%%%%%%%%%%%%%%%%%%%%%%%%%%%%%%%%%%%%%%%%%%%
     %                                           %
     %              F I G U R E S                %
     %                                           %
     %%%%%%%%%%%%%%%%%%%%%%%%%%%%%%%%%%%%%%%%%%%%%

\textheight=24.2cm
\nwp

%  #01
\begin{figure}
\vs{6cm}

\epsfysize=11.7cm
\epsffile[20 150 200 500]{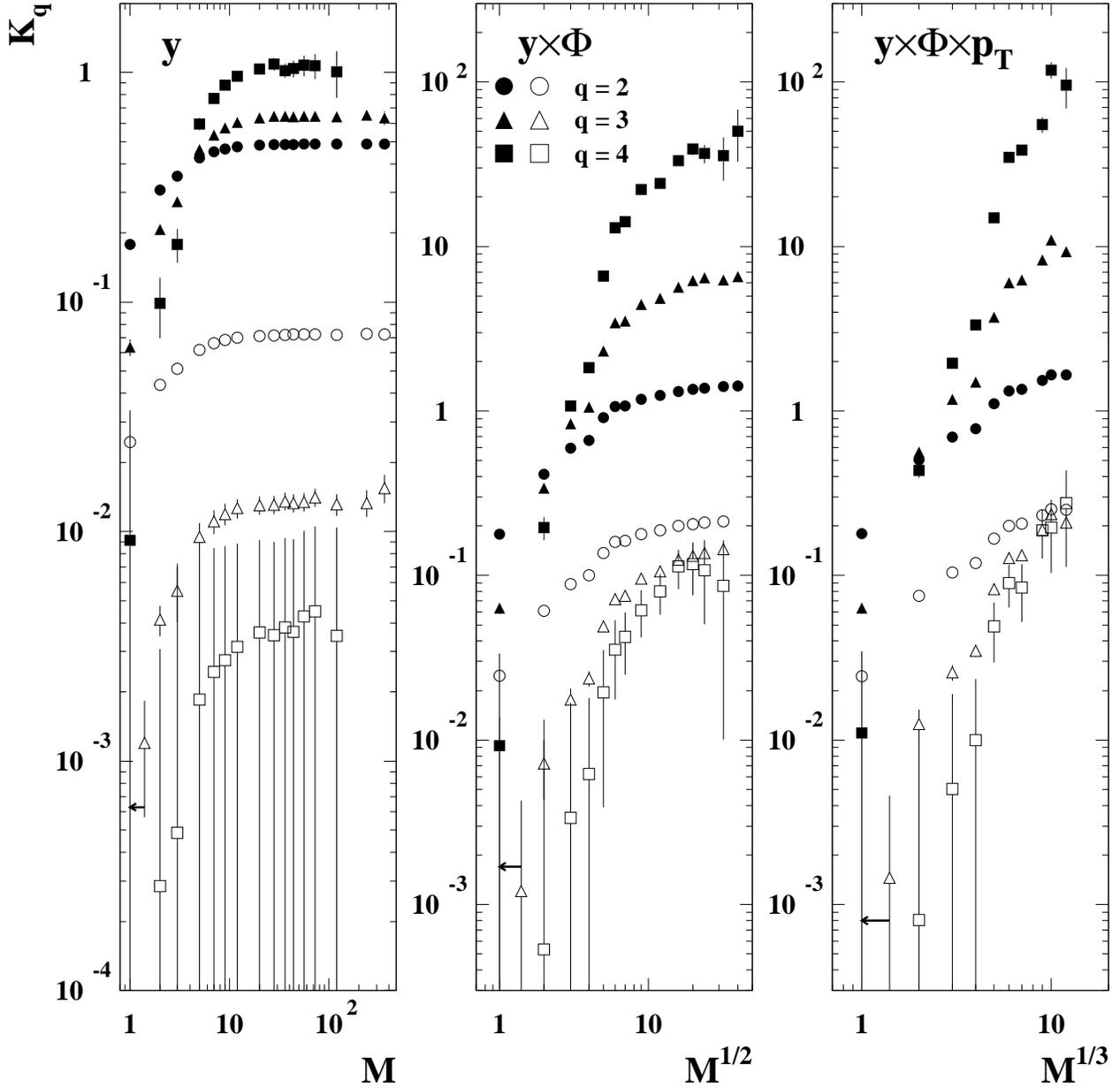}
\vs{1.cm}
\caption{\it 
The MC predicted cumulants of order $q =$ 2, 3 and 4 as a function of
$M^{1/D}$, where $M$ is the number of bins of the $D$-dimensional
subspaces of the phase space of rapidity ($y$), azimuthal angle ($\phi$),
and transverse momentum ($\pT$).
 The solid symbols represent the cumulants for a single source, while the
open symbols are the cumulants values of seven sources.
 }
\la{1f}
\end{figure}

\nwp 
% #02 
\begin{figure} 
\vs{6cm} 
\hs{3.2cm} 
\epsfysize=12.cm
\epsffile[45 150 200 500]{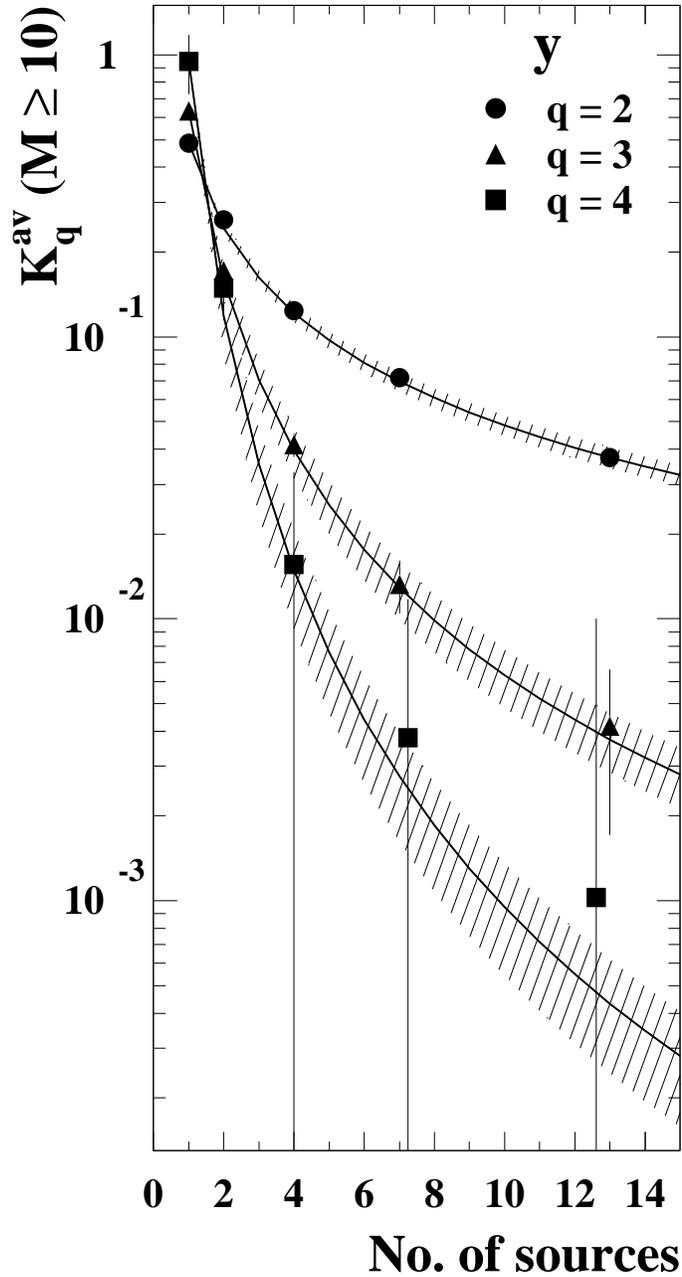} 
\vs{1.cm} 
\caption{\it 
 The dependence of the averaged rapidity 
cumulants $K_q^{\rm av}$ of order
$q=$ 2, 3 and 4 on the number of {\ep} sources. 
 The cumulants were averaged over the $M$-range where they approached a
constant value (see Fig. 1).
 The lines represent the expected dilution according to Eq. (\ref{difi})
where $q$ is neglected in comparison to the multiplicity $n$.
 The striped areas are the allowed regions when $q$ is not neglected with
respect to $n$ (see text).
 }
\la{1c}
\end{figure}

\end{document}